# Failover of Software Services
# with State Replication

**Karsten Wolke, K. Yermashov,**
**K. H. Siemsen, Rolf Andreas Rasenack,**

FH Oldenburg/Ostfriesland/Wilhelmshaven    De Montfort University, Software
Fachbereich Technik, INK    Technology Research Laboratory
Constantiaplatz 4    The Gateway
26723 Emden    Leicester LE1 9BH, UK

**ABSTRACT:** *Computing systems are becoming more and more complex and assuming more and more responsibilities in all sectors of human activity. Applications do not run locally on a single computer any more. A lot of today's applications are built as distributed system; with services on different computers communicating with each other. Distributed systems arise everywhere. The Internet is one of the best-known distributed systems and used by nearly everyone today. It is obvious that we are more and more dependant on computer services. Many people expect to be able to buy things like clothing or electronic equipment even at night on the Internet. Computers are expected to be operational and available 7 days a week, 24 hours a day. Downtime, even for maintenance, is no longer acceptable.*

## 1. Introduction

We will present initial work on a framework for the development of highly available software services. We focus on failover functionality. Failover is the migration of services from one server to another. If one server in the distributed system fails, another server takes over the service of the failed one. We believe that failover with state replication can increase the availability of services dramatically. In case of failure of one service instance, the clients continue their session on another service instance. In doing so, the clients automatically use the previous sequences and

267

operations, especially GUI inputs, without repeating them. The failover process is transparent to the client.

Today's development of available systems is very expensive, since every component of the system must be reliable. We believe that our framework can also decrease the costs to implement availability radically, so that even small business companies will be able to offer highly available services.

This article describes a concept how the functionality of failover can be implemented for services. A specific framework that provides the functionality of failover for Java services will be presented. The concept can also be adapted to services in other programming languages. The framework uses meta information to provide failover for services. Meta information and aspect oriented programming will be used in the framework to increase the efficiency of the work of developers. Developers only have to define some attributes to ensure failover with state replication. Code needed in software will be generated automatically by the framework.

## 2. The High Availability and Failover Framework

It is hard to find a company that is willing to accept longer server downtimes. Services in distributed systems, also on the Internet, have to be continuously available. If there is an IT-failure, it often implies considerable financial loss. Customers have little understanding for failures and simply change the provider. For this reason, high availability concepts are being increasingly applied in distributed systems, to decrease the risk of a failure of the computer systems. In reality 100 percent availability is not achievable [Wol06a]. There is no way to guarantee that a system is fault-free and 100 percent available. In fact, development of fault-free and 100 percent available systems involves infinite investments.

Therefore project leaders have to define an availability of less than 100 percent. The development becomes very expensive toward 100 percent system availability.

The High Availability and Failover Framework (HAFF) for Java Services came into being as an extension of an existing High Availability Framework (HAF). The HAF ensures that a client is always assigned to a running service. The functionality of failover ensures that the state of a server application (service) is saved at specific points (failover points). If a service-side failure occurs during the communication, the client does not repeat the calls of the failed session completely on a new application server, but will continue to run at approximately the position at which there was a failure on the previous application server.

The High Availability and Failover Framework (HAFF) [Wol03], shown in figure 1, offers services with high availability. The main parts of the HAFF are Observer, Administration-Server, Administration-Client and Failover-Management. Services are offered and administered by application servers. Observers monitor the application servers and the services running on them. Should a service or an application server fail, it is noticed by the observer and reported to the administration server. Clients will automatically be assigned to another application server that also provides the requested service. All failed services that were previously used by the clients will be started again and are therefore available. The Failover Management is used to guarantee that the clients can continue their session at approximately the position at which there was a failure in the previous application server. Without Failover management the clients have to repeat their interaction (e.g. user input).

The Administration-Server has information about the complete distributed system. All application servers and all offered services in the network are known by the Administration-Server. The data can be adjusted and viewed via an Administration-Client.

Figure 2 shows, for example, a screenshot of the Admin-Client of the reference implementation. It shows the availability of a Java service `FlightBooking` on application server `Apphope0`. Uptimes, downtimes and overload times are illustrated visually. Thus, administrators are able to see in an easy way when and where problems and failures occurred.

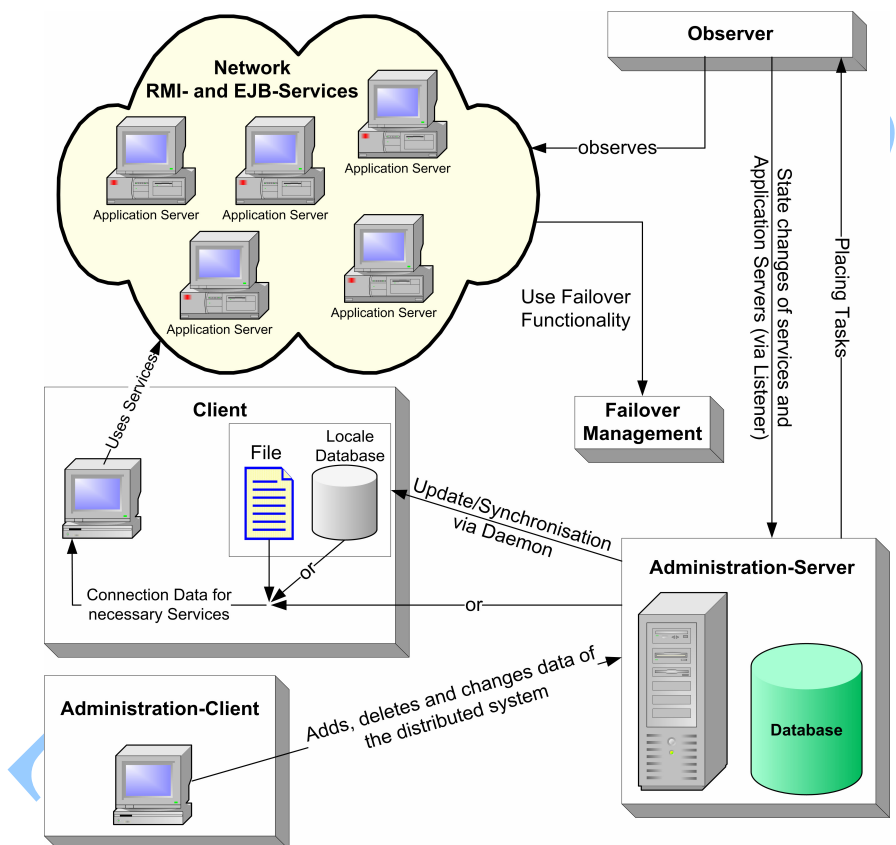

*Fig. 1:* *High Availability and Failover Framework*

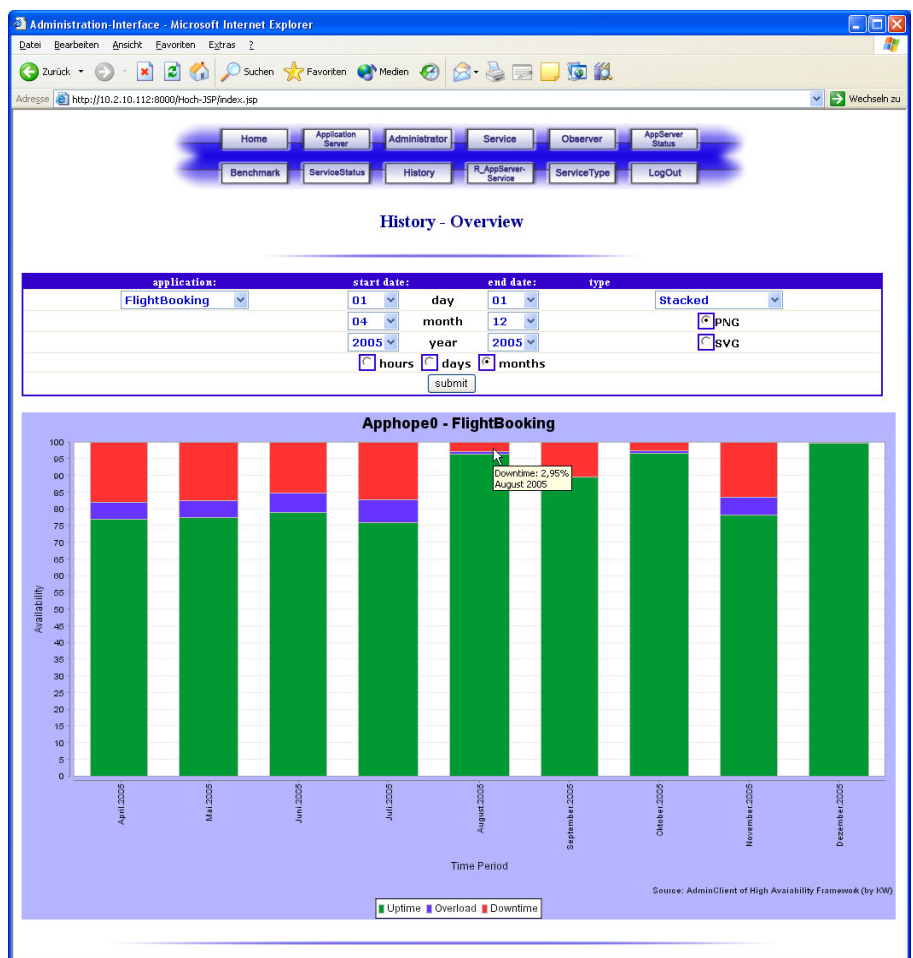

***Fig. 2:*** *Administration-Client, History of a Service*

In figure 1 the components of the HAFF are illustrated as logic components. All components of the HAFF are itself designed as services. This guarantees that the HAFF is high-available itself, since all components can be deployed redundantly and can be monitored by the observers.

As a consequence of the functionality of failover, the client continues the session at approximately the position at which there was a failure in the previous application server. This means the clients do not have to repeat the whole communication to the service.

This functionality is implemented in such a way that a developer of a service specifies so called failover points at any desired places in the source code. At each failover point the status will be saved persistent in a backend.

In case of a failure the status up to the last failover point (FOP) is recovered. The service then continues to run. The failover process is transparent to the client. A client will usually not notice a failure. A client will be automatically assigned to a new server that provides the requested services.

The FOPs are defined as multi-line comments (meta-information), so that if required, the service can run without the functionality of failover. If failover is desired, the developer starts a process that manipulates the source code automatically. This procedure examines the source code for FOPs and replaces the meta-information by failover code. A wizard can help the developer in the definition of FOPs.

## 3. Architecture of the Failover Framework

The source code of a service will be manipulated in an automatic way to apply failover functionality to it. These manipulations replace the FOP (meta-information, comments) by failover functionality. Code to load and store the states and controlling logic is generated automatically. Since it is tedious to carry out such manipulations at the source code level, an Abstract Syntax Language Tree (ASLT) [W+04] is used to simplify this process. The ASLT enables a view and manipulation of the source code in a tree form. Structures can be generated and modified via an ASLT-API in a simple way. Figure 3 shows that a "CodeToASLT" process converts the source text into an ASLT object. The FOP comments are saved as meta-information in the tree [Wol06b]. In a further process a pre-processor analyzes the FOPs and other meta-information defined in the ASLT object. If required, nodes are modified, moved or new nodes are generated. Finally the ASLT object is converted with an "ASLTToCode" process into source code files. The complete transformation process is illustrated in figure 3.

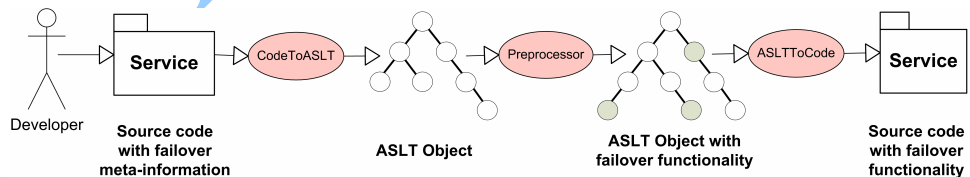

***Fig. 3:*** *Transformation of Failover Meta Information*

## 4. Transformation of Failover Meta Information

The example in Listing 1 demonstrates how failover meta-information (MI) in Java source codes looks like. The structure of meta-information in source code is defined in [Wol06b].

```
/*<MISet>
failover.VarSetDef{"VarSet1",str,i}
failover.VarSetDef{"VarSet2",i,j,k}
</MISet>*/

public void test() {

... //Code before failover point

/*<MISet>
 failover.Failoverpoint{j,#VarSet1}
</MISet>*/
/* MetaInfStatement */

...// Code after failover point
}
```

*Listing 1, Source code with Meta-information*

Before the method `test()`, there are two variable sets defined with the names `VarSet1` and `VarSet2`. Within the method `test()` a FOP is defined. There are Java statements before and after it. The FOP specifies, that the variable `j` and the variable set `VarSet1` (variables `i` and `str`) reflect the state of this position. The method can be continued at this point with the values of these variables.

After the generation of the ASLT object, it is passed on to the pre-processor. The meta-information for failover will be replaced by the corresponding constructs ("if-else", calls of the Failover Framework etc.). Listing 2 shows the modified code of listing 1 in a simplified form. To ensure failover, a unique ID is assigned to every failover point, which internally consists of 4 values (MethodID, MethodFOPCount, LevelCount, SessionID). Each method has a unique MethodID. The FOPs within a method are numbered in serial order by MethodFOPCount. A session of a client is identified by a SessionID. In order to ensure failover with nested method calls and recursion, the value LevelCount is used, which is incremented/decremented at runtime. Figure 4 demonstrates the meaning of these four IDs on an example scenario. In the example a FOP is defined before every method call. The methods `a()` and `c()` thus have two FOPs each.

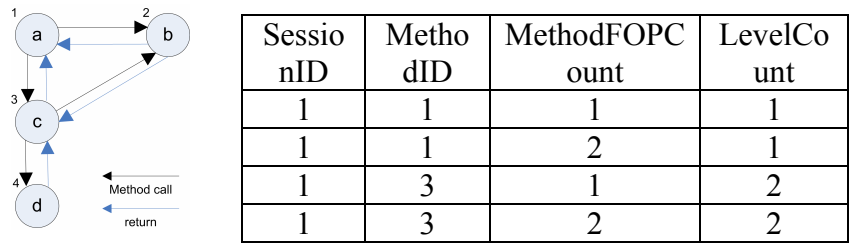

| SessionID | MethodID | MethodFOPCount | LevelCount |
|---|---|---|---|
| 1 | 1 | 1 | 1 |
| 1 | 1 | 2 | 1 |
| 1 | 3 | 1 | 2 |
| 1 | 3 | 2 | 2 |

*Fig. 4: Unique identification of Failover points*

In Listing 2, it can be seen that the pre-processor generates several lines of code. For the sake of clarity, sequences of statements are replaced by comments. At the start and end of the method, the LevelCount is incrementted or decremented. An if-else statement is generated for

```
public void test() {
levelCount++
if(fo.getFOPCount(sID,mID,levelCount) < 1) {
  // no FOP reached
  ... //Code before failover point
  ... //Store state of  j,i,str
}else if(fo.getFP() == 1) {
  //FOP 1 reached
  ... //Recover state of j,i,str
}
... // Code after failover point
levelCount--
}
```

*Listing 2, Source code with Failover code*

every FOP. The code before the FOP is placed into the "then" part. At the end of the "then" part, the state is stored persistent. The condition of the if-else -statement determines whether the FOP has already been reached once.

The object `fo` represents the connection to the Failover Management. The method `getFOPCount()` returns the MethodFOPCount of the FOPs that has been last passed within an execution block. If the result is `0`, no FOP has been passed. That is the case, for example, when no server-side failure has occurred. However, should there be a failure of the server after the FOP, `getFOPCount()` returns a `1` after the restart of the service. In this case, the code before the FOP is not executed once again. The state of the service is recovered at the FOP and the execution will be continued from there.

## 5. Failover Management

The Failover Management (FOM) provides the administration of the FOPs and the failover functionality. The functions of the Failover Management are defined in the interface `FailoverManagement` (see figure 5).

Several FOM implementations can thus be provided (for example, one FOM can use files for the persistence of the states; another FOM uses a database). A service gets a specific instance of a FOM via a factory [G+01] (`FailoverFactory`). An XML-file is used to control which service is handled by which FOM.

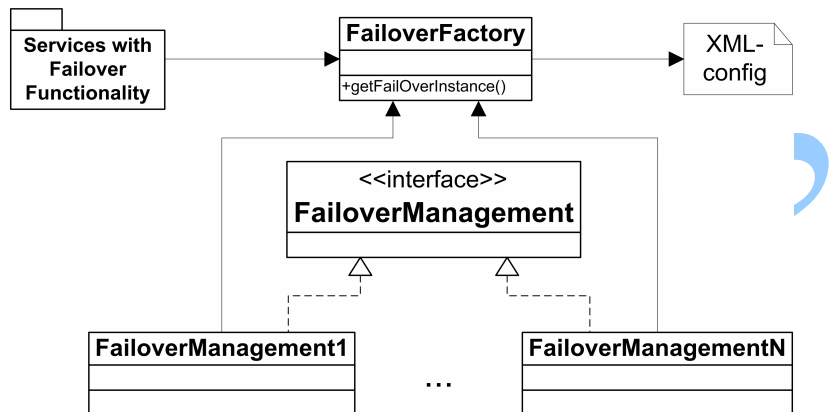

*Fig. 5: Failover Management*

**Summary**

The project presented here is not completed. It is extended continuously. From the theoretical point of view, the concepts of failover presented here are complete. To substantiate the theoretical concept, an example is implemented, which shows the functionality in the cases recursion and nested method calls. The current version of the failover frameworks can be downloaded under [W+06]. Since the implementation is not complete at the present time, there is still no information available about the performance of the concept. It is assumed that in case of a failure and restart of a time-intensive application, this concept will provide significant runtime advantages, especially in case of time-consuming calculation routines. In any case, it will be useful in session-based GUI applications, especially Web applications (e.g. online shops). No user likes it to input his painstakingly compiled shopping cart more than once. The framework decreases the costs to implement availability radically, so even small business companies will be able to offer highly available services that are operational and available 7 days a week, 24 hours a day. Meta information (multi line comments) are

275

defined by developers to ensure failover with state replication. The framework generates the code for failover functionality automatically. The reference implementation demonstrates the power of the proposed framework for Java Services. The concept can be adapted to services in other programming languages.

**References**

[G+01]     Gamma E., Helm R., Johnson R., Vlissides J. - *Design Patterns, Elements of Reusable Object-Oriented Software*, Addison Wesley. 2001

[Wol03]    Wolke K. - *RMI and Enterprise Java Beans, High Availability and Extended Functionality*, Thesis at the Fachhochschule Ostfriesland, http://www.karsten-wolke.de/Diplomarbeit/doc/ Diplomarbeit.pdf, 2003.

[Wol06a]   Wolke K. - *Transfer Report, Reliability of distributed Enterprise Systems, High Availability, Fail-Safety and Failover Functionality*, STRL, DeMontfort University Leicester, http://www.karsten-wolke.de/public/phd/Transfer.pdf, 2006

[Wol06b]   Wolke K. - *Meta Information and its Processing*, Fachhochschule Oldenburg/ Ostfriesland/ Wilhelmshaven, Standort Emden, Fachbereich Technik and STRL, DeMontfort University Leicester, http://www.karsten-wolke.de/public/aslt/ ASLTMetaData.pdf, January 2006

[W+04]     Wolke K. et all - *Abstract Syntax Trees for Source Code Management*, Toolbox magazine, September/October 2004

[W+06]     Wolke K. et all - *Failover Framework*, Fachhochschule Oldenburg/Ostfriesland/Wilhelmshaven, Standort Emden, Fachbereich Technik and STRL, DeMontfort University Leicester, http://www.karsten-wolke.de/public/aslt/ ASLT_1.1.rar, 2006